%% file: main.tex
\documentclass[sigconf,screen]{acmart}

\usepackage[braket, qm]{qcircuit}

\usepackage{physics}
\usepackage{nicefrac}
\usepackage{bbold}
\usepackage{graphicx}
\usepackage[normalem,normalem,normalbf]{ulem}
\usepackage{tabu}
\usepackage{subfig}
\usepackage{balance}

\setcopyright{acmcopyright}
\acmPrice{15.00}
\acmDOI{10.1145/3575693.3575726}
\acmYear{2023}
\copyrightyear{2023}
\acmSubmissionID{asplosb23main-p453-p}
\acmISBN{978-1-4503-9916-6/23/03}
\acmConference[ASPLOS '23]{Proceedings of the 28th ACM International Conference on Architectural Support for Programming Languages and Operating Systems, Volume 2}{March 25--29, 2023}{Vancouver, BC, Canada}
\acmBooktitle{Proceedings of the 28th ACM International Conference on Architectural Support for Programming Languages and Operating Systems, Volume 2 (ASPLOS '23), March 25--29, 2023, Vancouver, BC, Canada}
\received{2022-07-07}
\received[accepted]{2022-09-22}

\begin{document}

\title[Qompress: Efficient Compilation for Ququarts]{
Qompress: Efficient Compilation for Ququarts Exploiting Partial and Mixed Radix Operations for Communication Reduction}

\author{Andrew Litteken}
\email{litteken@uchicago.edu}
\affiliation{%
  \institution{University of Chicago}
  \city{Chicago}
  \state{IL}
  \country{USA}
}

\author{Lennart Maximilian Seifert}
\email{lmseifert@uchicago.edu}
\affiliation{%
  \institution{University of Chicago}
  \city{Chicago}
  \state{IL}
  \country{USA}
}

\author{Jason Chadwick}
\email{jchadwick@uchicago.edu}
\affiliation{%
  \institution{University of Chicago}
  \city{Chicago}
  \state{IL}
  \country{USA}
}

\author{Natalia Nottingham}
\email{nottingham@uchicago.edu}
\affiliation{%
  \institution{University of Chicago}
  \city{Chicago}
  \state{IL}
  \country{USA}
}

\author{Frederic T. Chong}
\email{chong@cs.uchicago.edu}
\affiliation{%
  \institution{University of Chicago}
  \city{Chicago}
  \state{IL}
  \country{USA}
}

\author{Jonathan M. Baker}
\email{jmbaker@uchicago.edu}
\affiliation{%
  \institution{University of Chicago}
  \city{Chicago}
  \state{IL}
  \country{USA}
}

\renewcommand{\shortauthors}{A. Litteken, M. Seifert, J. Chadwick, N. Nottingham, F. Chong, J. Baker}



\date{}
\input{paper-text/00-abstract}
\maketitle

\thispagestyle{empty}

\input{paper-text/01-introduction}

\input{paper-text/02-background}
\input{paper-text/03-pulse-generation}
\input{paper-text/05-qudit-compilation}
\input{paper-text/06-compression}
\input{paper-text/07-evaluation-methods}
\input{paper-text/08-results}
\input{paper-text/09-conclusion}

\begin{acks}
This work is funded in part by EPiQC, an NSF Expedition in Computing, under award CCF-1730449; in part by STAQ under award NSF Phy-1818914; in part by NSF award 2110860; in part by the US Department of Energy Office of Advanced Scientific Computing Research, Accelerated Research for Quantum Computing Program; and in part by the NSF Quantum Leap Challenge Institute for Hybrid Quantum Architectures and Networks (NSF Award 2016136) and in part based upon work supported by the 
U.S. Department of Energy, Office of Science, National Quantum Information Science Research Centers.  FTC is Chief Scientist for Quantum Software at ColdQuanta and an advisor to Quantum Circuits, Inc.

We would like to thank Casey Duckering for his input in early discussion of compiler development for ququarts. We would like to thank Stefanie Günther and N. Anders Petersson for valuable advice on using the quantum optimal control software packages Juqbox and Quandary. Additionally, we would like to thank David I. Schuster for helpful discussions regarding quantum optimal control theory.

This work was completed in part with resources provided by the University of Chicago’s Research Computing Center.
\end{acks}

\bibliographystyle{ACM-Reference-Format}
\balance
\bibliography{references_latest}

\end{document}

%% file: paper-text/00-abstract.tex
\begin{abstract}
Quantum computing is in an era of limited resources. Current hardware lacks high fidelity gates, long coherence times, and the number of computational units required to perform meaningful computation. Contemporary quantum devices typically use a binary system, where each qubit exists in a superposition of the $\ket{0}$ and $\ket{1}$ states. However, it is often possible to access the $\ket{2}$ or even $\ket{3}$ states in the same physical unit  by manipulating the system in different ways. In this work, we consider automatically encoding two qubits into one four-state qu\emph{quart} via a \emph{compression scheme}. We use quantum optimal control to design efficient proof-of-concept gates that fully replicate standard qubit computation on these encoded qubits.

We extend qubit compilation schemes to efficiently route qubits on an arbitrary mixed-radix system consisting of both qubits and ququarts, reducing communication and minimizing excess circuit execution time introduced by longer-duration ququart gates. In conjunction with these compilation strategies, we introduce several methods to find beneficial compressions, reducing circuit error due to computation and communication by up to 50\%.  These methods can increase the computational space available on a limited near-term machine by up to 2x while maintaining circuit fidelity.\end{abstract}

%% file: paper-text/01-introduction.tex
\begin{figure}[H]
    \centering
    \vspace{-1.0em}
    \includegraphics[width=0.78\linewidth]{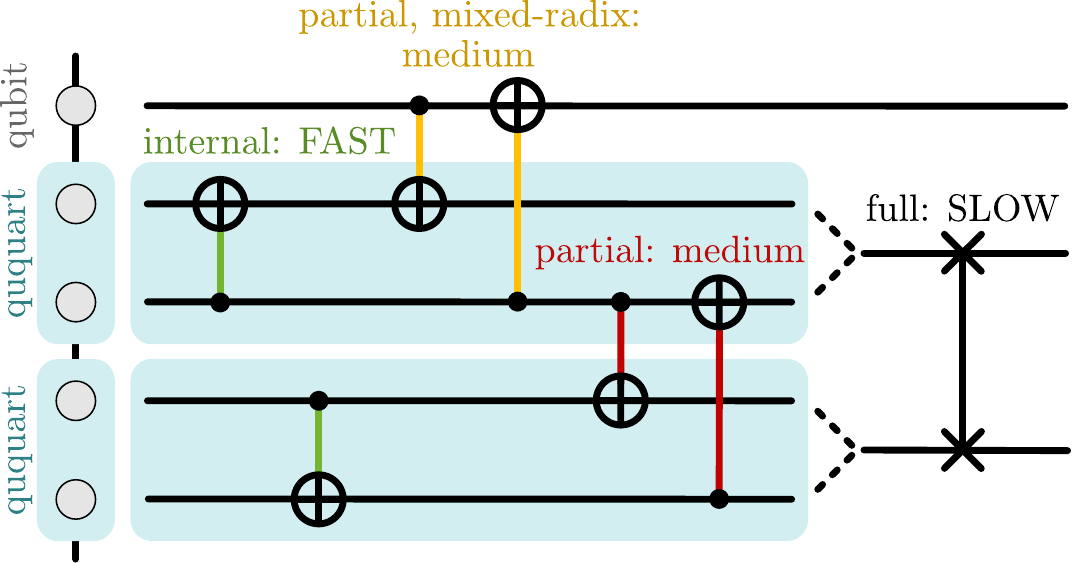}
    \caption{Pairs of qubits can be compressed in four-dimensional ququarts and interact with each other internally or through partial operations, enabling novel compilation techniques and space reduction.}
    \label{fig:intro}
    \vspace{-1.0em}
\end{figure}

\section{Introduction}
Quantum computing has the potential to change the way we think about what is computationally tractable.  Today, quantum systems exist, but there are far fewer qubits than what are required for large error-corrected algorithms \cite{shor_polynomial-time_1997, grover_fast_1996}. There are product design roadmaps to take us into the realm of thousands of qubits \cite{gambetta_expanding_2022}, but quantum programmers will still be pushing against hardware space constraints, looking for optimizations to maximize hardware utilization.

A typical quantum architecture is constructed of binary qu\textit{bits}, which have two distinct states representing 0 and 1. However, this binary abstraction is not the whole picture. Some quantum hardware, such as superconducting qubits \cite{cho_tunable_2015} and trapped ions \cite{ringbauer_universal_2021}, have natural access to energy levels beyond the 0 and the 1 states that can be used as  extra computation space. Previous studies have adapted qubit-based quantum programs to temporarily access these higher states in order to improve circuit fidelity and circuit runtime \cite{baker_improved_2020, litteken_communication_2022, krishna_generalization_2016}. However, these studies have been focused on hand optimization of specific subcircuits (such as the generalized Toffoli gate) and applications instead of providing a general approach for any quantum circuit.

Alternatively, higher qudit states can be used in a more general way to save computational space. The information of two qubits can be fully \emph{encoded} in a single ququart \cite{baker_efficient_2020}, storing two logical units' worth of information in a single computational unit of higher radix. However, this strategy has been avoided due to several drawbacks that scale with qudit dimension: quadratically increasing logic gate execution time \cite{lloyd_efficient_2019}, reduced coherence time, and increased difficulty of experimentally applying gates \cite{houck_life_2009, blok_quantum_2021}.  These are significant disadvantages for a NISQ device that already has limited connectivity, short qubit lifetimes, and high gate error. However, if these issues can be further mitigated, this general compression method can effectively \textit{double} the available computational space on a quantum computer for any arbitrary circuit.

Previous work has attempted temporary ququart encoding to harness the potential for fast "internal" gates between the two encoded qubits, but had to decode for any operation outside of a specific ququart. In this work, we use quantum optimal control to synthesize a specific set of gates, that can interact two qubits in the same ququart, a qubit outside of a ququart with a qubit encoded in a ququart, or two different qubits encoded in two different ququarts ququarts, allowing us to selectively compress certain qubit pairs while leaving others as bare qubits. Each of these gates have varying speeds, and are shown in Figure \ref{fig:intro}.

A mixed-radix paradigm where qubits are directly manipulated within an encoded ququart adds a degree of flexibility not found in the prior work. Inter-ququart operations no longer incur the extra cost of encoding and decoding. Motivated by these new gates, we explore strategies to efficiently generalize partial ququart compression of a circuit to be used with any qubit-based circuit. The partial qubit-ququart and mixed-radix operations inform the design of a compilation pipeline designed to take advantage of the flexibility of our gate set while minimizing the error due to decoherence.

We propose several different strategies to find the ideal compression to make the best use of the newly-found ququart compilation methods.  The main contributions of this work are the following:
\begin{itemize}
    \item Develop a library of high-fidelity mixed qubit-ququart operations for partially compressed qubit systems via quantum optimal control for a realistic device Hamiltonian.
    \item Detail a compiler pipeline that takes advantage of partial ququart operations for mixed-radix computing.
    \item Evaluate several compression strategies to selectively encode qubits in ququarts using the new compiler pipeline.
    \item Demonstrate simulated increases in gate fidelity of more than 50\% over standard qubit-only compilation across various quantum architectures, as well as up to 2x increased qubit capacity, while accounting for the increased coherence errors associated with operation on ququarts.
\end{itemize}

%% file: paper-text/02-background.tex
\section{Background}

\subsection{Quantum Computing}
The fundamental unit of quantum computation is the qubit. Unlike a classical bit, a qubit can exist in a linear superposition, of $\ket{0}$ and $\ket{1}$ as $\ket{\psi} = \alpha\ket{0} + \beta\ket{1}$.  A quantum program, or circuit, acts on $N$ qubits, and its state can be represented as a superposition of  $2^N$ bitstrings. Logic gates manipulate this superposition; for example, the X operation flips the $\ket{0}$ and $\ket{1}$ states of a single qubit, performing the operation $\mathrm{X} \ket{\psi} = \beta\ket{0} + \alpha\ket{1}$. 

Multi-qubit gates, such as CNOT, can be used to create entangling relationships between qubits.  Together, superposition and entanglement enable quantum computing to solve  problems that are potentially intractable on classical computers.


\subsection{Higher Radix Computation}
While the qu\textit{bit} abstraction focuses on only the lowest two energy levels of a quantum system as logical states, we can more generally consider qu\textit{dits} which utilize the lowest $d-1$ levels as logical states. Many hardware technologies have access to these energy levels, but are usually limited by the ability to effectively control them. Current systems typically only use these additional states to speed up the implementation of multi-qubit gates that still \textit{logically} use only two levels. This work focuses on qudits that occupy the lowest four energy levels ($\ket{0}$, $\ket{1}$, $\ket{2}$, and $\ket{3}$), which we call \textit{ququarts}. In general, qudit computation is not asymptotically better than qubit computation -- both schemes can universally express quantum computation, and full translation from one radix to another affords only constant advantages \cite{gokhale_asymptotic_2019}.  

Practical use of qudits (particularly qutrits and ququarts) at the application level have focused on hand optimization, often constrained to a small number of applications. These works \cite{baker_efficient_2020, litteken_communication_2022} focus primarily on temporary access of these logical states, observing primarily a reduction in space requirements by small expansions of the computation space of logical units such as temporarily accessing the third state to use as a carry bit. This prior work also demonstrates that an explicit \textit{compression} of qubits into ququarts can free up additional ancillary space to expand the usefulness of limited hardware, but this approach is only effective when circuits contain qubits explicitly known to be in the $\ket{0}$ state.

These works opt to avoid any computation on the compressed information, apart from compression and decompression, as it becomes increasingly challenging to effectively control qudits with larger dimension \cite{chi_programmable_2022}. Characterizing higher energy levels is difficult as they are more prone to noise and suffer from lower coherence times, at a rate of around $T_1/(d-1)$, where $d$ is the qudit dimension and $T_1$ is the coherence time for a qubit \cite{blok_quantum_2021}. So, every use of these higher levels is more prone to failure, especially for longer circuits, making them more difficult to efficiently utilize.


\subsection{Quantum Optimal Control}
Gates on a device are implemented by applying hardware-specific control fields $f_k(t)$ to the qudit(s) involved in the operation. In superconducting architectures, control fields are analog microwave pulses. Quantum optimal control searches for a control sequence that best replicates the effect of a target logic gate. Different algorithms and toolboxes have been designed for this purpose \cite{khaneja_optimal_2005, sklarz_loading_2002, petersson_optimal_2021, gunther_quantum_2021}. In this work, we use the open-source optimal control software Juqbox \cite{petersson_discrete_2020, petersson_optimal_2021} to find the shortest-duration control pulse sequence that reaches a specified fidelity for each logic gate of interest. Juqbox optimizes the control fields $f_k(t)$ to minimize an objective function $J[f_k] = 1 - F[f_k] + L[f_k]$ consisting mainly of the gate fidelity
\begin{equation}
    F[f_k] = \frac{1}{h^2} \abs{\Tr{U^\dagger_T[f_k] \, V}}^2
\end{equation}
between the target unitary $V$ and the applied transformation $U_T[f_k]$, where $h$ is the Hilbert space dimension of the logical subspace. This is achieved by repeatedly solving the Schrödinger equation and adjusting the control fields in every iteration to minimize $J$. The full Hilbert space of the optimization typically includes additional guard states to capture the influence of higher energy levels present in superconducting systems \cite{koch_charge-insensitive_2007, petersson_optimal_2021}. As these guard states are not part of the logical subspace, their populations are penalized with a leakage term $L[f_k]$ in the objective function.

\subsection{Related Work}
Use of additional logical levels for quantum computation is not new. There have been two primary considerations for their use. First, full translations from qubits to qudits, usually of small dimension $d$, have been proposed. For example, translation has been used to implement arithmetic or Shor's algorithm \cite{bocharov_factoring_2017, bocharov_improved_2016}, which require an expanded gate set to implement generalized ternary gates. Second, temporary use of additional logical states has been shown to be useful in specific cases. In these cases, programs begin and end entirely as qubits, but during computation temporarily access additional logical levels. These works focus on a small set of applications such as the generalized Toffoli gate and adders \cite{baker_improved_2020, wang_improved_2011} and rely on hand optimization to extract benefit from these states without using too many gates or spending too much time occupying these states. This temporary use strategy has been generalized as \textit{compression} \cite{baker_efficient_2020} to generate ancilla to speed up specialized subcircuits. These prior works operate in the gate model only, assuming that gates on different dimension qudits are effectively the same - they can be executed with equivalent fidelity and execute in a similar amount of time. They also generally ignore architectural connectivity constraints and the inherent increased cost to communicate qudits at long distances. While gate representations are useful, they omit crucial systems level details, such as gate duration and pulse implementation fidelity, which determine the viability of mixed radix computation in general. Prior work has shown a worst-case quadratic increase in gate \textit{duration} \cite{lloyd_efficient_2019} for higher-dimensional gates, meaning in practice we must be extremely careful about how we use qudits.

%% file: paper-text/03-pulse-generation.tex
\section{Qudit Pulse Generation}

Qudit logic gates are typically assumed to require long durations to implement, due to the increased complexity of the Hilbert spaces involved \cite{lloyd_efficient_2019}. This is a problem for several reasons: First, longer-duration gates lead to overall longer-duration circuits; second, higher qudit states have shorter decoherence times \cite{blok_quantum_2021}, amplifying the negative effect of these longer circuits. However, by explicitly synthesizing control pulses for higher-radix gates, we find gate durations that scale efficiently enough to provide overall circuit benefits when combined with a tailored compiler.

\subsection{Compression and Gate Set} \label{sec:gates}
We follow a qubit to ququart compression scheme inspired by \cite{baker_efficient_2020}. Encoding the state of two qubits $\ket{q_0 q_1}$ into a single ququart state promotes one of the qubits $q_0$ to a ququart while transforming the other qubit $q_1$ to an ancilla in the ground state $\ket{0}$. While this is referred to as a compression, we do not lose any data. It is the full representation of two qubits, in a single physical unit. We define the encoding gate ENC as
\begin{equation} \label{eq:enc}
    \mathrm{ENC} = \, \begin{cases} \, \ket{0}\ket{0} \rightarrow \ket{0}\ket{0} \\
    \, \ket{0}\ket{1} \rightarrow \ket{1}\ket{0} \\
    \, \ket{1}\ket{0} \rightarrow \ket{2}\ket{0} \\
    \, \ket{1}\ket{1} \rightarrow \ket{3}\ket{0}
    \end{cases}
\end{equation}
This gate defines the encoding scheme; for example, the $\ket{2}$ ququart state represents the $\ket{10}$ qubit-qubit state, and we can perform logical operations on the ququart states as if they are the states of two qubits. We note that, since we do not expect $q_0$ or $q_1$ to be in a ququart state prior to the encoding operation, the extension of ENC to a full ququart-ququart unitary gate is arbitrary.

Additionally, to measure a ququart, we simply use the mapping in reverse to determine the state of the two encoded qubits from the state of the ququart. In a physically realized system, we are not able to measure one encoded qubit without simultaneously measuring the other, unless the qubits are first decoded.

\subsubsection{Gate Set}
\begin{figure}[tbp]
    \centering
    \includegraphics[width=0.8\linewidth]{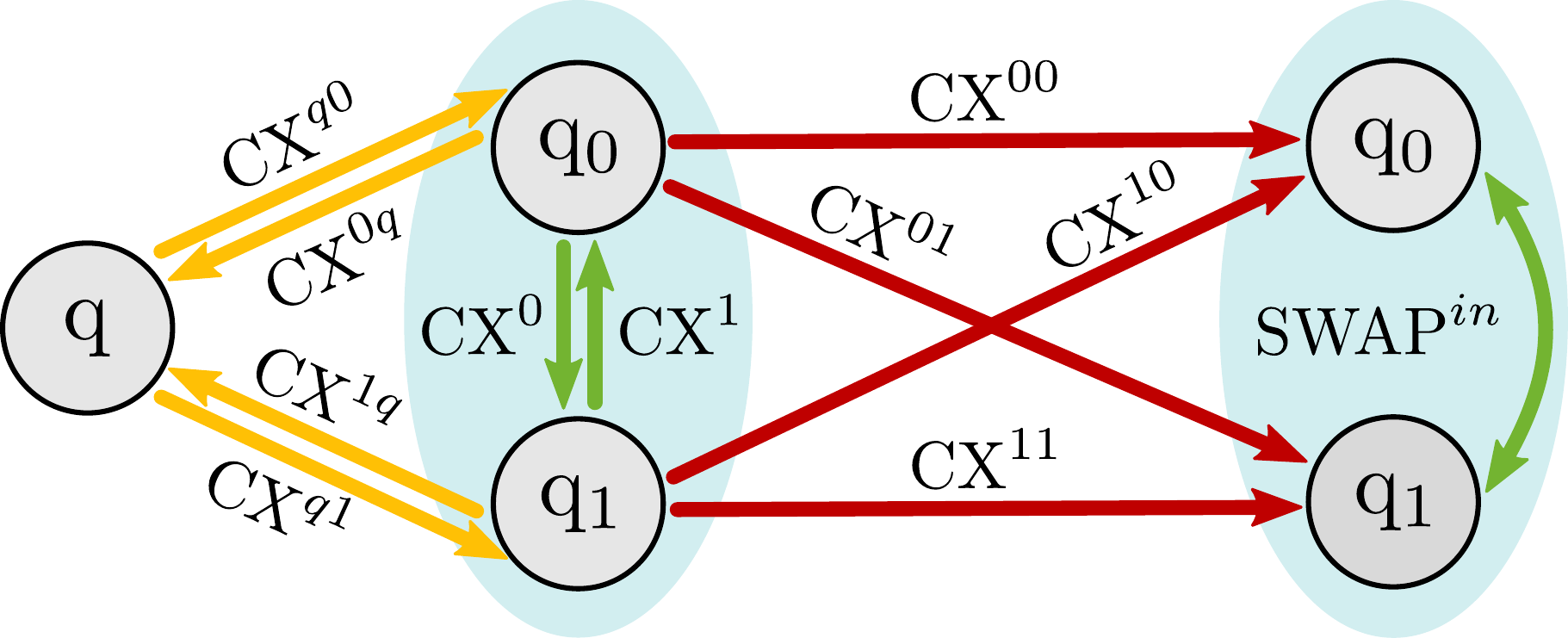}%
    \caption{Two qubits $q_0$ and $q_1$ can be encoded into a ququart (blue oval) and interact with each other internally (green), with a bare qubit $q$ outside (yellow), or with encoded qubits in a different ququart (red). CX arrows point from control qubit to target qubit. Along each CX link, corresponding SWAP gates are defined with the same superscripts/subscripts (not shown).}
    \label{fig:qudit_expansion}
\end{figure}

Under this encoding scheme, qubit-equivalent operations can be derived for ququarts that encode two qubits, some of which are shown in Figure \ref{fig:qudit_expansion}. 
For example, X gates acting on the encoded qubit state $\ket{q_0 q_1}$ correspond to the ququart operators $\mathrm{X}^0 \sim \mathrm{X} \otimes \mathbb{1}$ (which switches the population of $\ket{0}$ with $\ket{2}$, as well as $\ket{1}$ with $\ket{3})$ and $\mathrm{X}^1 \sim  \mathbb{1} \otimes \mathrm{X}$.
These gates can also be executed in parallel by applying a ququart gate $\mathrm{X}^{0,1} \sim \mathrm{X} \otimes \mathrm{X}$.

Any two-qubit gate acting on qubits encoded in the same ququart can be expressed as a single-ququart gate. For example, an \emph{internal} SWAP gate (SWAP$^{in}$) corresponds to a simple X$_{12}$ operation that exchanges populations of the ququart $\ket{1}$ and $\ket{2}$ states. Similarly, we can define internal CX gates $\{$CX$^0$, CX$^1\}$ between the two encoded qubits. On some hardwares, these operations can be done much faster (and with higher fidelity) than the corresponding two-qubit operations.
However, in order to compile quantum circuits for a mixed-radix quantum computer using both qubits and ququarts, two-qubit gates between a bare qubit and an encoded qubit as well as between encoded qubits in different ququarts are necessary. To this end we extend our gate set by a variety of \emph{partial} CX- and SWAP-like gates acting on various qubit pairs as shown in Figure \ref{fig:qudit_expansion}. We define four partial CX gates $\{$CX$^{q0}$, CX$^{q1}$, CX$^{0q}$, CX$^{1q}\}$ and two partial SWAP gates $\{$SWAP$^{q0}$, SWAP$^{q1}\}$ that realize the respective two-qubit gate between a bare qubit and an encoded qubit. Similarly, four partial CX gates $\{$CX$^{00}$, CX$^{01}$, CX$^{10}$, CX$^{11}\}$ and three partial SWAP gates $\{$SWAP$^{00}$, SWAP$^{01}$, SWAP$^{11}\}$\footnote{We focus only on unique gates; SWAP$^{01}$ and SWAP$^{10}$ are equivalent.} can manipulate the states of encoded qubits in different ququarts. We further include the full ququart-ququart SWAP$_4$ gate to enable routing of ququarts.


\subsection{Device Model}
We obtain durations of gates in our system by explicitly optimizing high-fidelity control pulses for superconducting hardware. We model two-qudit subsystems with two weakly coupled, anharmonic transmons \cite{koch_charge-insensitive_2007} with total Hamiltonian
\begin{equation}
\begin{aligned}
    H(t) = &~\sum_{k=1}^2 \qty[\omega_k a_k^\dagger a_k + \frac{\xi_k}{2} a_k^\dagger a_k^\dagger a_k a_k] + J \qty(a_1^\dagger a_2 + a_2^\dagger a_1) \\
    &+ \sum_{k=1}^2 f_k(t) \qty(a_k + a_k^\dagger),
\end{aligned}
\label{eq:ham_rot}
\end{equation}
where the first two terms comprise the drift Hamiltonian and the last describes the effect of the control fields $f_k(t)$ applied to the transmons.
We choose realistic physical parameters inspired by \cite{sheldon_procedure_2016}: The 0-1 transition frequencies of the transmons are $\omega_1/2\pi = 4.914 \,\mathrm{GHz}$ and $\omega_2/2\pi = 5.114 \,\mathrm{GHz}$, and both transmons have the same anharmonicity $\xi_1/2\pi = \xi_2/2\pi = -330 \,\mathrm{MHz}$. They are effectively coupled with $J/2\pi = 3.8\,\mathrm{MHz}$. For single-qudit gates we reduce this model to $k=1$ and remove the $J$ coupling term. We restrict the maximum amplitude of the control fields to $f_\mathrm{max} = 45\,\mathrm{MHz}$ to constrain potential leakage into guard states.


\subsection{Optimizing Pulses}

We aim to find control pulses of shortest \textit{duration} for each gate. As real quantum systems are limited by noise effects such as decoherence, this allows us to investigate the expected cost of using these gates in quantum circuits. Juqbox only allows pulse optimization for a fixed time interval $[0, T]$, therefore we minimize pulse durations $T$ by applying the technique from \cite{seifert_time-efficient_2022}, which involves iterative re-optimization with previous pulse results. This method aims to find the shortest-duration pulse that can execute the desired gate above a minimum fidelity target. In this work, our target fidelity for single-qudit gates is $F=0.999$ and for two-qudit gates $F=0.99$.

\begin{figure}[tbp]
    \centering
    \includegraphics[width=\linewidth]{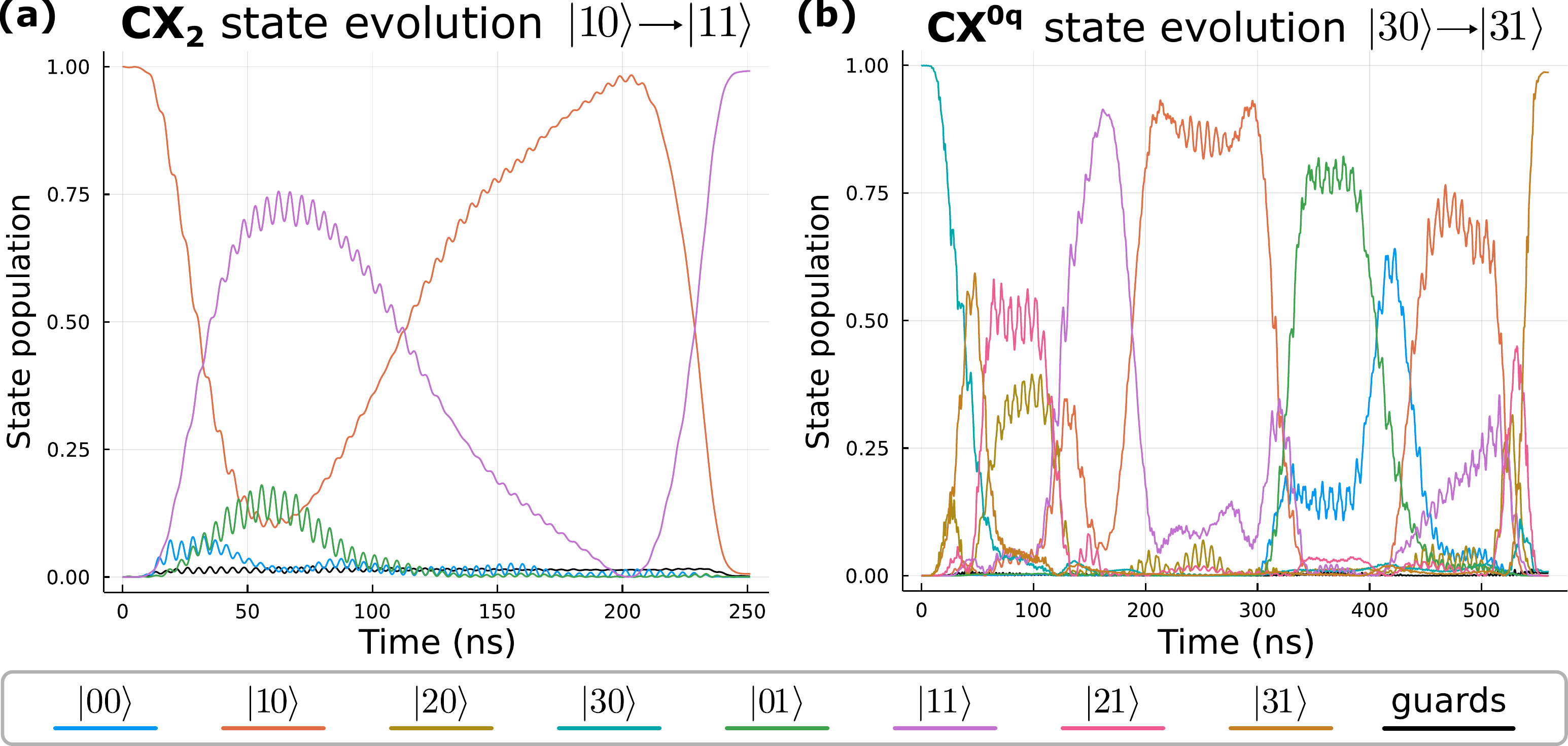}
    \caption{Exemplary state evolutions of two CX gates between (a) two bare qubits and (b) a bare qubit and an encoded qubit. (a)  
    The control qubit $q_0$ is in state $\ket{1}$, hence the state of the target qubit $q_1$ is flipped. (b) The encoded qubit $q_0$ inside the ququart controls the CX gate targeting the bare qubit $q$. The ququart state $\ket{3}$ corresponds to the two-qubit state $\ket{q_0 q_1} = \ket{11}$, so the state of $q$ is flipped.}
    \label{fig:cx_evolutions}
\end{figure}

Figure \ref{fig:cx_evolutions} visualizes equivalent state evolutions in two different CX gates: the standard CX$_2$ between two bare qubits and the partial CX$^{0q}$ controlled by an encoded qubit and targeting an unencoded qubit. In both cases, the control qubit is in state $\ket{1}$ (the ququart state $\ket{3}$ is equivalent to the encoded qubit state $\ket{11}$), causing the target state to flip from $\ket{0}$ to $\ket{1}$. An important observation is the difference in complexity of the state dynamics for the two gates. CX$^{0q}$ operates on twice as many states in the logical subspace as CX$_2$. This suggests that the difficulty of finding high-fidelity control pulses increases with Hilbert space dimension and explains the variation in gate durations we observe when repeatedly optimizing for complex two-qudit gates.


\subsection{Qudit Gate Durations}
\label{sec:gate_durations}
Table \ref{tab:gates} shows the minimum-duration results for logic gates of interest (described in Section \ref{sec:gates}). In the compiler and benchmarks, we assume that any single-qubit gate will have a similar duration to an X gate, so we do not explicitly find pulses for other single-qubit gates such as H or Z. SWAP gates, which move data around the architecture, are extremely common in limited connectivity devices; while they can be decomposed into three CNOTs, we find that optimized pulses can perform a SWAP operation in far shorter time than this would imply, so we explicitly optimize SWAP gates.

\begin{table*}[htbp]
    \caption{Shortest pulse durations found for gates of interest (see Figure \ref{fig:qudit_expansion}), in nanoseconds. a) CNOT and SWAP operations between two encoded qubits in the same ququart become single-ququart operations, making them significantly faster. b) Standard two-qubit gates for comparison. c) Gates between one encoded qubit and one bare qubit. (d) Gates between two encoded qubits in different ququarts. Note: CX$^{10}$ and CX$^{11}$ minimum durations converged to larger values than shown; however, these can be done faster using internal SWAP operations and CX$^{00}$, giving a duration of 78 + 544 + 78 = 700 ns.}
    \renewcommand{\arraystretch}{1.2}
    \begin{tabu}{l r|l r|[2pt]l r|[2pt]l r|l r|[2pt]l r|l r}
        \multicolumn{4}{c|[2pt]}{\textbf{(a) Qudit}} & \multicolumn{2}{c|[2pt]}{\textbf{(b) Qubit-Qubit}} & \multicolumn{4}{c|[2pt]}{\textbf{(c) Qubit-Ququart}} & \multicolumn{4}{c}{\textbf{(d) Ququart-Ququart}}\\
        \hline
        
        X & 35 & X$^0$ & 87 & CX$_2$ & 251 & CX$^{0q}$ & 560 & CX$^{q0}$ & 880 & CX$^{00}$ & 544 & CX$^{01}$ & 544 \\

        X$^1$ & 66 & X$^{0,1}$ & 86  & SWAP$_2$ & 504 & CX$^{1q}$ & 632 & CX$^{q1}$ & 812 & CX$^{10}$ & 700 & CX$^{11}$ & 700 \\

        CX$^0$ & 83 & CX$^1$ & 84 &&& SWAP$^{q0}$ & 680 & SWAP$^{q1}$ & 792 & SWAP$^{00}$ & 916 & SWAP$^{01}$ & 892 \\

        SWAP$^{in}$ & 78 &&&&& ENC & 608 &&& SWAP$^{11}$ & 964 & SWAP$_4$ & 1184 \\
    \end{tabu}
    \label{tab:gates}
\end{table*}

We observe several interesting relationships across gate types. Using an internal CNOT or SWAP instead of its corresponding qubit-qubit operation gives a significant speedup. Additionally, a SWAP between a bare qubit and an encoded qubit (680 or 792 ns) is significantly faster than a SWAP between two encoded qubits (892-964 ns). We emphasize that these relationships are specific to this Hamiltonian, and another system may have different tradeoffs in gate durations. Our goal is to design a compiler that can adapt to different sets of gate durations to determine favorable ququart encodings, without explicitly depending on a specific relationship such as the internal CNOT advantage shown here.

\subsection{Technical Barriers}
The main technical barrier to a physical implementation of this scheme is accurate control of quantum logical units at higher energy levels, not the ability to access them. There are several documented machines that use qutrits \cite{blok_quantum_2021, hill_beyond_2021, cervera-lierta_experimental_2022, galda_implementing_2021} and demonstrations of ququart devices \cite{chi_programmable_2022}; although calibration of these devices is more time-consuming, it is not a fundamental limitation. In superconducting systems, as the energy level increases, increased charge noise causes phase errors and faster state decay, making control more difficult \cite{koch_charge-insensitive_2007, houck_life_2009, blok_quantum_2021}.

There is precedent for use of more than two energy levels, e.g. as protection or guard states or to improve measurement and expedite qubit-level gates \cite{koch_charge-insensitive_2007, krishna_generalization_2016, geerlings_demonstrating_2013, lucero_high-fidelity_2008}, indicating that it is possible to reliably use higher-energy states.  However, at present, we are not aware of experiments using highly-optimized pulses for these energy levels, which would be a potential challenge for the results shown in this work.

As access to calibrated devices with high-dimensional qudits is extremely limited, our methods are currently difficult to physically verify, and the efficiency of controlling these energy states remains an open question. The overhead and difficulty of controlling higher energy states are potential issues and may change the benefits found in the work. We attempt to be as realistic as possible by designing this scheme with a transmon Hamiltonian based on IBM’s real hardware, similar to that which has experimentally demonstrated support for qutrit operations \cite{galda_implementing_2021}. Quantum optimal control has experimentally been proven successful in synthesizing accurate pulses within a certain error threshold for qubit-only gates on similar devices \cite{abdelhafez_universal_2020}, so we are optimistic that the scheme and compilation strategy laid out in this work will be applicable. Additionally, in an ongoing collaboration with the Schuster group \cite{schusterlab}, 
we are working to validate our methodology and have to date demonstrated some single ququart 
gates with high fidelity. While the control pulses that we obtained in this section do not correspond to a real, 
fully characterized device, we again emphasize that the compilation techniques discussed in the following section are independent of the specific control pulses used, and will adapt to gate durations and error rates different than obtained here. We intend this work to act as a proof-of-concept to demonstrate the potential value of accessing higher-energy qudit states and as a motivating factor to improve ququart characterization and control.


%% file: paper-text/05-qudit-compilation.tex
\section{Qompress: Compilation with Partial Operations}
\label{sec:compilation}
Compiling a circuit for an architecture which supports full encoding of qubits into ququarts is not fundamentally different from compiling in a qubit-only architecture.  However, we must account for higher connectivity and varying efficiency of operations between different pairs of qubits based on their configuration.  This includes a new library of communication gates, including the encoding gate, mixed-radix gates, and cross-ququart gates.  We extend current compiler technologies to account for these new variables and fully utilize the unique set of gates to minimize space requirements while increasing expected circuit success rate and ideally minimizing increased circuit durations.

\subsection{Representing a Ququart Architecture}
Each quantum unit in the system can either be treated as a bare qubit only accessing the lowest two energy levels, or a full ququart which can access the lowest four. We represent this mixed-radix architecture via two graphs.  One represents the overall topology of the architecture, where each node represents a single quantum logic unit and edges represent allowed communication between these units.  The second represents the logical qubits we will be mapping onto from our original circuit.  In this graph, each physical unit is treated as if it is a ququart, and is expanded into two qubit nodes that are connected to one another, called an interaction graph.  Both qubit nodes in the expanded ququart are connected to each qubit node contained in adjacent ququarts. For example, if a ququart was connected to $n$ other ququarts, each encoded qubit is connected to $2n + 1$ other encoded qubits.  In the secondary qubit graph, there are $2V$ nodes and $4E + V$ edges, where $V$ and $E$ are the number of physical logic units and number of interacting pairs of units, respectively. This expansion is shown in Figure \ref{fig:qudit_expansion}. Edges are annotated with a set of execution times and error rates for CX gates and SWAP gates, which differ depending on whether either unit is a bare qubit or encoded within a ququart.

\subsection{Extending Mapping, Routing, and Scheduling}
Current quantum architectures do not have all-to-all connectivity; in limited-connectivity architectures, non-adjacent operands require communication to be inserted into the circuit, adding costs such as increased gate count and/or duration. In general, the goal of mapping, routing, and scheduling is to reduce these costs as much as possible, as both contribute to reduced circuit success. Mapping and routing has been studied extensively in the qubit-only case \cite{tannu_ensemble_2019, hirata_efficient_2011, cowtan_qubit_2019} and in some initial studies for intermediate qudits \cite{litteken_communication_2022}.

However, in our proposed mixed-radix system, not all communication is equivalent, or even similar, in cost. A CX gate between a qubit and ququart is lower fidelity and has a longer duration than a CX gate between encoded qubits inside the same ququart. Therefore, our compilation schemes aim to reduce error and circuit duration by taking advantage of gates unique to ququart systems.

The goal of mapping is to place frequently-interacting qubits close to one another on the device.  Since the ququart architecture is modeled as a set of qubits with different connections, we can easily extend mapping strategies designed for qubit-only architectures to ququart-based architectures. First, we find an interaction weight between each pair of qubits in the original circuit.  We use the weighting function $w(i, j) = \sum_{o \in C} \mathbb{1}(i, j \in o_q)/s(o)$, where $o$ is an operation in circuit $C$ to represent the interactions between qubits $i$ and $j$, and $s$ is a function from the operation $o$ to an integer time step in the circuit, starting from 1.  Denoting $Q_c$ the set of qubits in the circuit, we select the qubit that maximizes the highest total weight to other qubits $W(i) = \sum_{j \in Q_c \setminus \qty{i}} w(i, j)$.  We find the center-most ququart in the architecture and place this qubit in the first encoding position. For each unmapped qubit, we compute the greatest sum of weights to the already placed qubits and score each potential placement by how strongly it interacts with mapped qubits and its distance from them.

In our mixed-radix system, we only consider the second encoding location in ququarts if the first encoding location has already been mapped to. For qubits only, distance can be a simple shortest path calculation, with edges weighted by the fidelity of the connection. We do the same for ququarts, but with dynamic weighting based on the current encoding of qubits.  We represent the current duration of gate $g$ at connection $(i, j)$ as $T(i, j, g)$ and the fidelity as $F(i, j, g)$.  So, for a gate at a given connection, the probability of success is $S(i, j, g) = F(i, j, g)e^{-T(i, j, g)/T_{1,i}}e^{-T(i, j, g)/T_{1,j}}$, where $T_1$ is the decoherence time for a qubit or a ququart, whichever is being used. This is a common metric of success for a quantum gate as used in \cite{stein_eqc_2022}. The aggregate probability of success for a given path $P_n$ of length $n$ is then modeled as:
\begin{equation}
\begin{aligned}
    S(P) = -&\log(S(P_{n-1}, P_{n}, \mathrm{CX})) \\
         +& \sum_{i=1}^{n-2} \qty[-\log(S(P_i, P_{i+1})), \mathrm{SWAP}))]\\
\end{aligned}
\label{eq:prob_of_success}
\end{equation}

This calculation allows us to account for errors introduced by both the decoherence time and the number of gates. We repeat this process iteratively until every program qubit has been mapped to a hardware location represented in the extended ququart graph.

In our system, qubits are still tracked individually even if they are both stored in the same physical object. Therefore, routing for a mixed-radix device is still performed at the qubit level. Any qubit routing strategy, such as lookahead strategies \cite{wille_look-ahead_2016, jandura_improving_2018, anis_qiskit_2021}, can be translated directly to encoded qubits in ququarts through routing based on the logical qubit architecture graph rather than the ququart-level graph. The main goal is to disrupt the current mapping as little as possible as qubits that need to interact are moved closer to one another. We choose the candidate location which disrupts the current state of the circuit the least and moves the qubits closer together based on the probability of success laid out in Equation \eqref{eq:prob_of_success}.

However, we do place some constraints on qubit movement.  First, we do not encode new ququarts during routing. Second, we avoid swapping ``through'' ququarts when possible.  Ququart operations remain expensive, and pairings determined during mapping are often beneficial to reducing execution costs. Both of these will incur extra costs.  Additionally, by placing these restrictions on the compiler, we ensure that there are fewer changes in the maximum energy level of the computational units, we can cache the calculated distances, significantly reducing the amount of classical computation required.

Circuits using fully encoded ququarts require more serialization than an analogous qubit-only circuit.  If, at a given time, each qubit in a fully encoded ququart is involved in a CX operation, we can no longer execute both gates in parallel and they must be sequenced one after the other. To break this tie and avoid bottlenecks, we select whichever operation is on the longest execution path of remaining qubits.  This can also be an issue for two single-qubit gates targeting two qubits in the same encoded ququart.  In this case, we combine both operations into one single-ququart gate, as executing one gate acting on a full ququart is less error prone than executing two single-qubit gates.

%% file: paper-text/06-compression.tex
\section{Qubit Compression Strategies} \label{sec:strategies}
Poor initial mapping can incur high communication costs later on. Additionally, ququart gates are inherently more time-intensive and can incur serialization, increasing circuit duration and error due to decoherence.  Simply extending qubit mapping strategies to the expanded ququart-based graphs will not take any of these factors into account, or take full advantage of the enhanced connectivity of fully encoded ququarts. In this section, we explore various compression strategies to better account for these features.

\subsection{Exhaustive Compression (EC)}
The true effect of individual compression circuit duration or fidelity cannot be reliably predicted without recompiling the circuit with a prescribed compression. Further, finding the optimal set of compressions is computationally difficult on its own since there are exponentially many possible subsets. To get a sense for an upper bound on circuit quality, we explore an exhaustive, but iterative and greedy, search of qubit compressions.

\begin{figure}[tbp]
    \centering
    \includegraphics[width=0.7\linewidth]{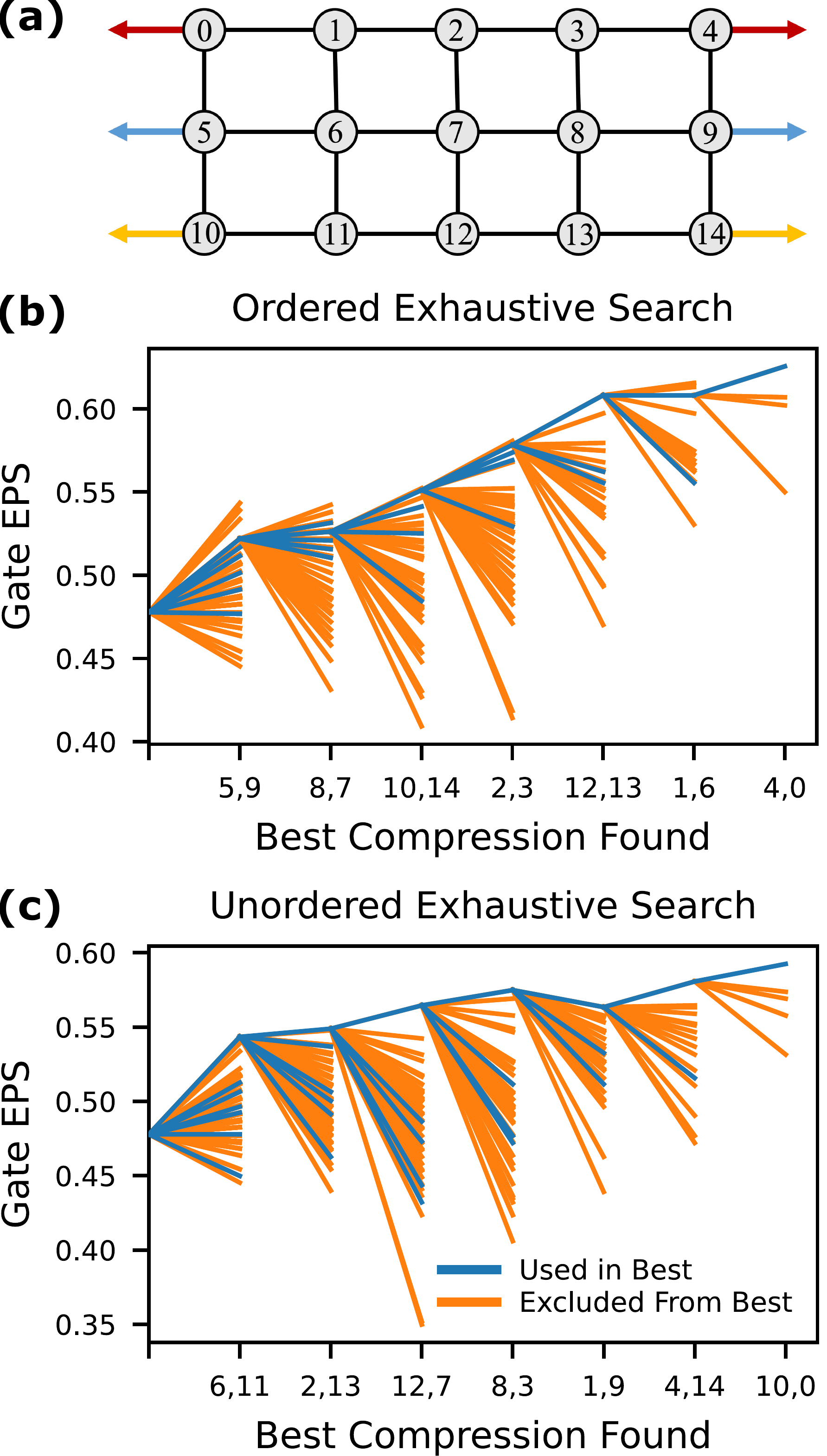}
    \caption{An example of an exhaustive search on a cylinder-based interaction graph (a), using a (b) critical path focused selection, and a (c) selection strategy that allows for any pair to be selected.}
    \label{fig:exhaustive-examples}
\end{figure}

At each step, we recompile the input circuit with every pair of qubits compressed and choose the option that maximizes the circuit fidelity. Searching every pair, while more complete, is computationally demanding. Circuit duration is defined by the length of the critical path and therefore it may be more advantageous to explore only compressions which affect the critical path. When choosing compressions we prioritize in order compressions which affect: 1) qubits in non-communication gates on the critical path, 2) qubits off the critical path that insert communication operations along the critical path, and 3) qubits off the critical path. In group order, we choose the best compression, if any, and repeat. We compare this ordered selection to an unordered selection in Figure \ref{fig:exhaustive-examples}. 

As can be seen in Figure \ref{fig:exhaustive-examples}, it is not the case that the compression of any two qubits will be advantageous to the combined gate success rate or error due to decoherence. There are two classes of advantageous compression.  Some compressions are beneficial due to high interactivity between the two encoded qubits, winning from the fast internal gates. An example is the pairing of qubits $6$ and $11$ in the circuit shown in Figure \ref{fig:exhaustive-examples}(a).  These qubits interact very often, so compression increases the probability of success of CX gate by turning it into a single-qudit gate. The second compression type wins by reducing communication, such as the pairing of qubits $4$ and $14$ in the cylinder graph based circuit.  These qubits rarely interact, but this reduces the \textit{diameter} of the circuit, decreasing required SWAP gates. Critical path prioritization and an unordered search of the entire space result in similar success rate gains, but different compressions but depends on the circuit structure.

Examining all potential compressions is quadratic in the number of qubits in the circuit, on top of the relatively high complexities required for mapping and routing. Using the insights found through this method, we build strategies that are able to approximate these benefits.

\subsection{Extended Qubit Mapping (EQM)}
EQM relies entirely on the interactions between the pairs of qubits, and follows the algorithm in Section \ref{sec:compilation}. There is no explicit pair selection in this method. The qubits are filled greedily, based on the highest weight to the already placed qubits. This strategy clusters qubits that interact often closely together, but will likely preemptively encode two qubits as it sees immediate benefits based on those already placed, focusing on identifying interaction-based compressions. But, it does have classical advantages: EQM is wrapped into the mapping step, and the only additional complexity is due to the expanded qudit graph.

\subsection{Ring Based (RB)}

\begin{figure*}[tbp]
    \centering
    \includegraphics[width=0.8\textwidth]{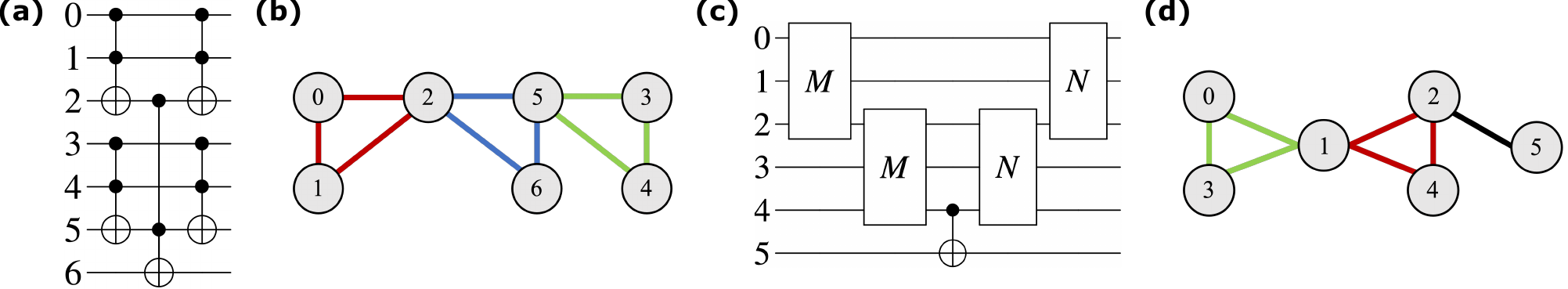}
    \caption{(a), (b) Sample generalized Toffoli gate with corresponding interaction graph. (c), (d) Cuccaro adder with interaction graph, M and N are three qubit gate blocks. Some circuits, such as the CNU circuit and Cuccaro adder have clusters of qubits that interact with one another. These clusters can be identified by finding the cycles in the interaction graph.  We have identified the cycles here with colored edges.}
    \label{fig:ring-examples}
\end{figure*}

While frequency of interaction is a good metric for identifying locality, it fails to account for more holistic interaction structures. Figure \ref{fig:ring-examples} shows that for certain circuits, such as the generalized Toffoli gate and the Cuccaro adder, there is a regular, triangle-based structure in the interaction graph.  We can use compressions of these triangles to transform the interaction graph into a line.  Looking at Figure \ref{fig:ring-examples}(b), this means compressing pairs 0 and 1, 2 and 6, and 3 and 4. A linear interaction graph is a much more favorable structure to mapping.  In fact, for certain configurations, it can be mapped and routed without any SWAPs on any architecture. 

We can generalize this triangle compression to any size cycle.  If we can find beneficial compressions within a given set of cycles, we may be able to further transform circuit interactions for more effective mappings.  
It is computationally expensive to find all cycles in an undirected graph, therefore for each qubit we find the minimum size cycle which includes it. This ensures that each qubit is contained in at least one cycle without finding all cycles.  We find the minimum cycle length from this set of cycles, and use it to bound the maximum identifiable cycle size.

After we identify cycles, we find the qubits with the fewest interactions outside their cycle and test compressions with each other qubit in the cycle.  For each compression, we collect information such like the number of shared neighbors between qubits, the weight of interaction between the compressed qubits, how often they are interacted on simultaneously, and how many cycles this pair appears in.

We prioritize compressions that maximize the number of internal interactions and the number of connections to other qubits while minimizing the time the encoded qubits are used at the same time, to make full use of both the higher fidelity internal ququart interactions and the new partial ququart operations without introducing serialization.  
We pick the best compression based on these metrics.  The qubit nodes are removed from the interaction graph and replaced with a single node representing the pair, with edges connected to each interacting qubit or pair. Following these adjustments to the interaction graph, we recollect interaction statistics and repeat for other pairings until no other beneficial compressions can be made.

\subsection{Average Weight per Edge (AWE)}
Another method, Average Weight per Edge, makes compressions to maximize the average edge weight of the interaction graph. This method takes advantage of shared interactions to increase potential locality and reducing communication.

We examine each pair of qubits and select the pair that maximizes the average weight per edge once the two qubit node is collapsed into a single node. We repeat until no more compressions can be made, or until there is no pairing that would increase the average weight per edge.

\subsection{Progressive Pairing (PP)}
The final strategy is an extension of EQM, which is more computationally intensive but collects more information.  We start by mapping the circuit to a qubit-only architecture.  This provides a full picture of how the circuit can be laid out on device, compared to the incremental version provided by EQM.  We again examine each potential pairing and compute the estimated fidelity with and without the compression based on changes in distance between interacting pairs without remapping and rerouting. In this scheme there are two choices when compressing two qubits A and B, either A is first or B is first. We select compression adjustment that is expected to increase the fidelity the most.

After selecting the pair that has the greatest increase in estimated overall fidelity, we remap and re-evaluate the circuit with the given pair and repeat the process. We continue this process until we cannot find any pair that will reduce the estimated overall fidelity. While this method scales quadratically, we avoid mapping and routing for each possible pairing by only determining distances.  While less scalable than the other methods, Progressive Pairing attempts to replicate the exhaustive search with much less classical overhead.

%% file: paper-text/07-evaluation-methods.tex
\section{Evaluation Methods}\label{sec:evaluation-methods} 
\begin{figure}
    \begin{minipage}[c]{0.3\linewidth}
    \centering
    \subfloat[Cylinder Graph]{
        \centering
        \includegraphics[width=\textwidth]{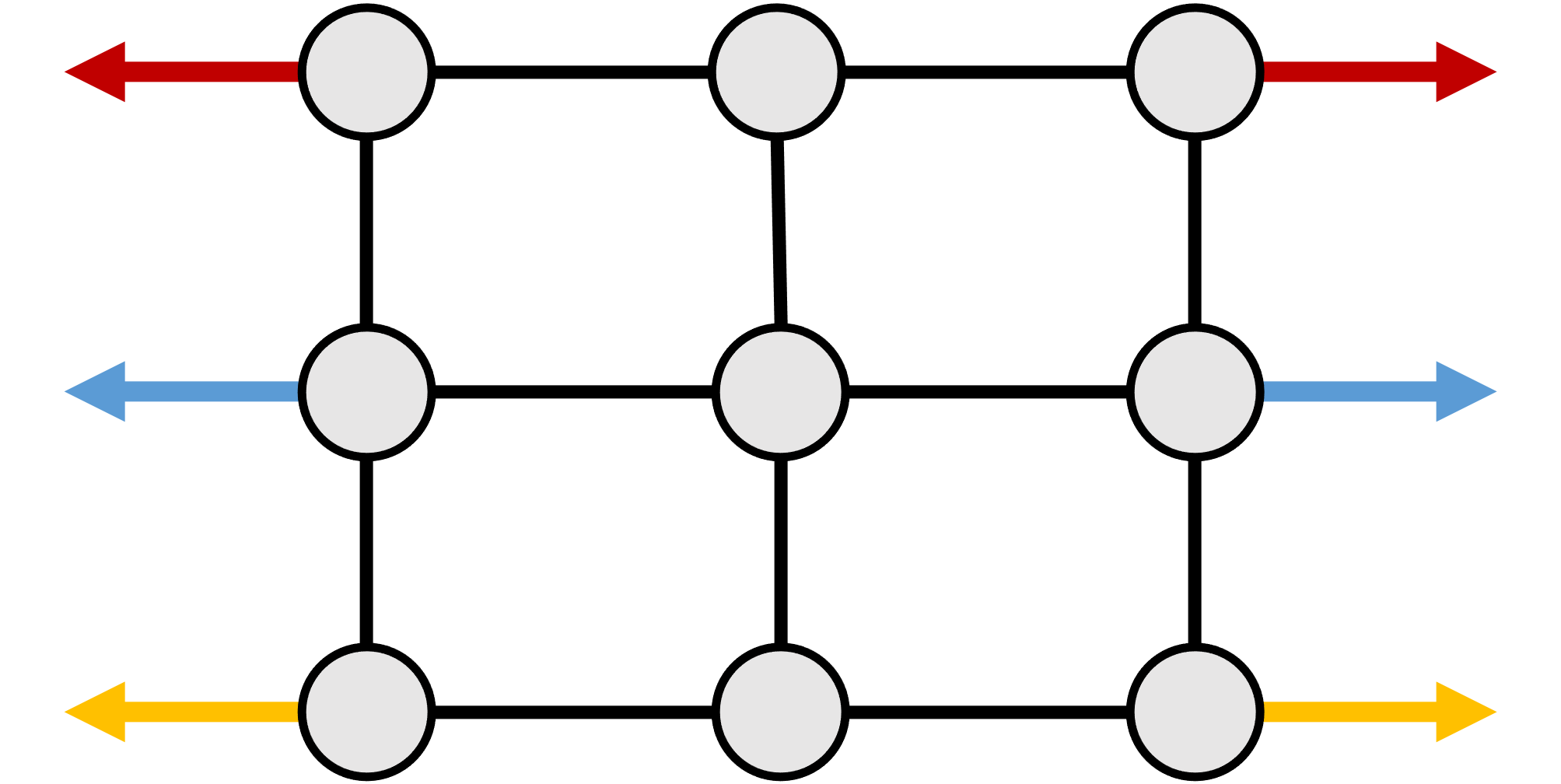}
    }
    \end{minipage}
    \begin{minipage}[c]{0.3\linewidth}
    \centering
    \subfloat[Torus Graph]{
        \centering
        \includegraphics[width=\textwidth]{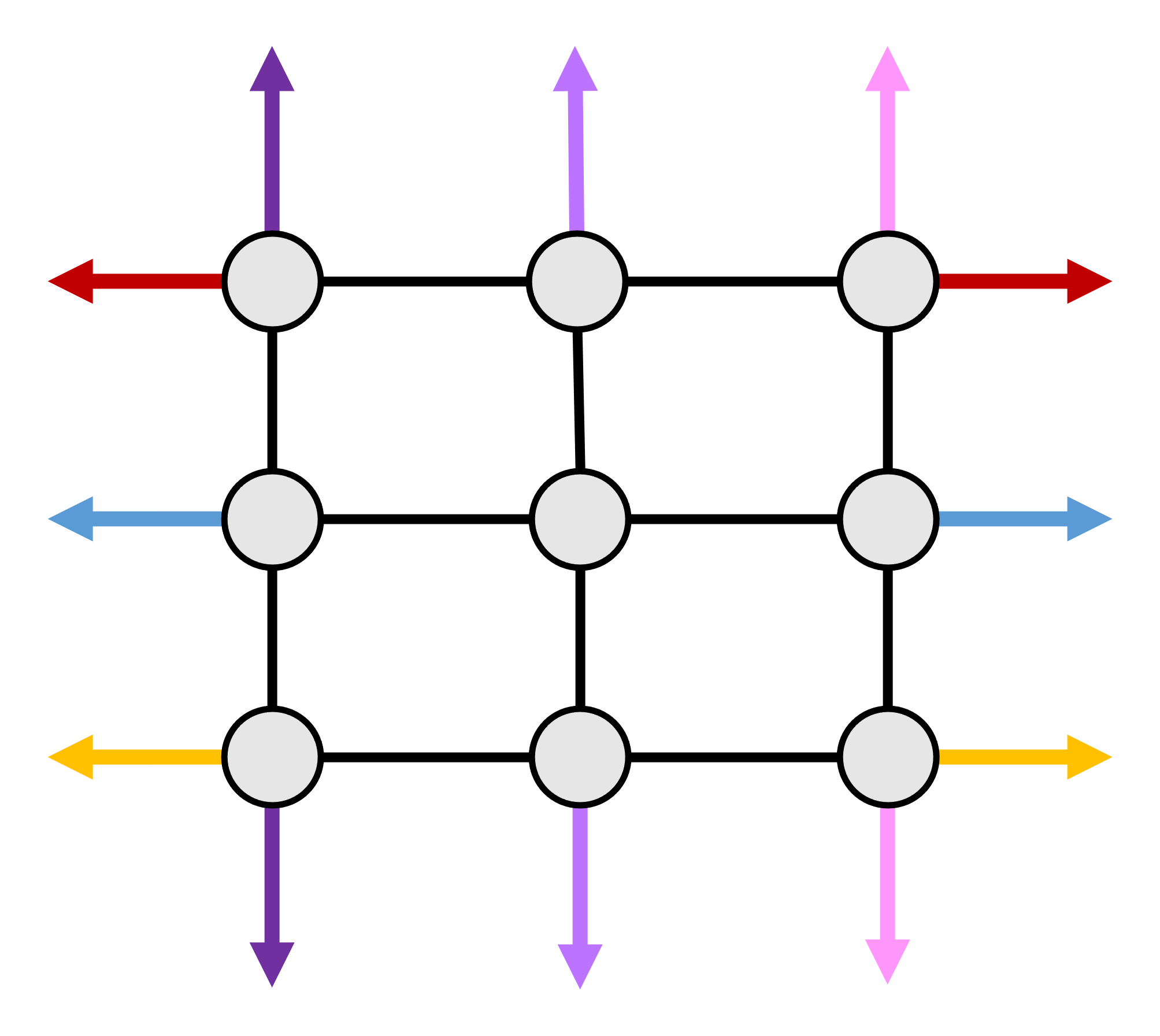}
    }
    \end{minipage}
    \begin{minipage}[c]{0.3\linewidth}
    \centering
    \subfloat[Binary Welded Tree]{
        \centering
        \includegraphics[width=\textwidth]{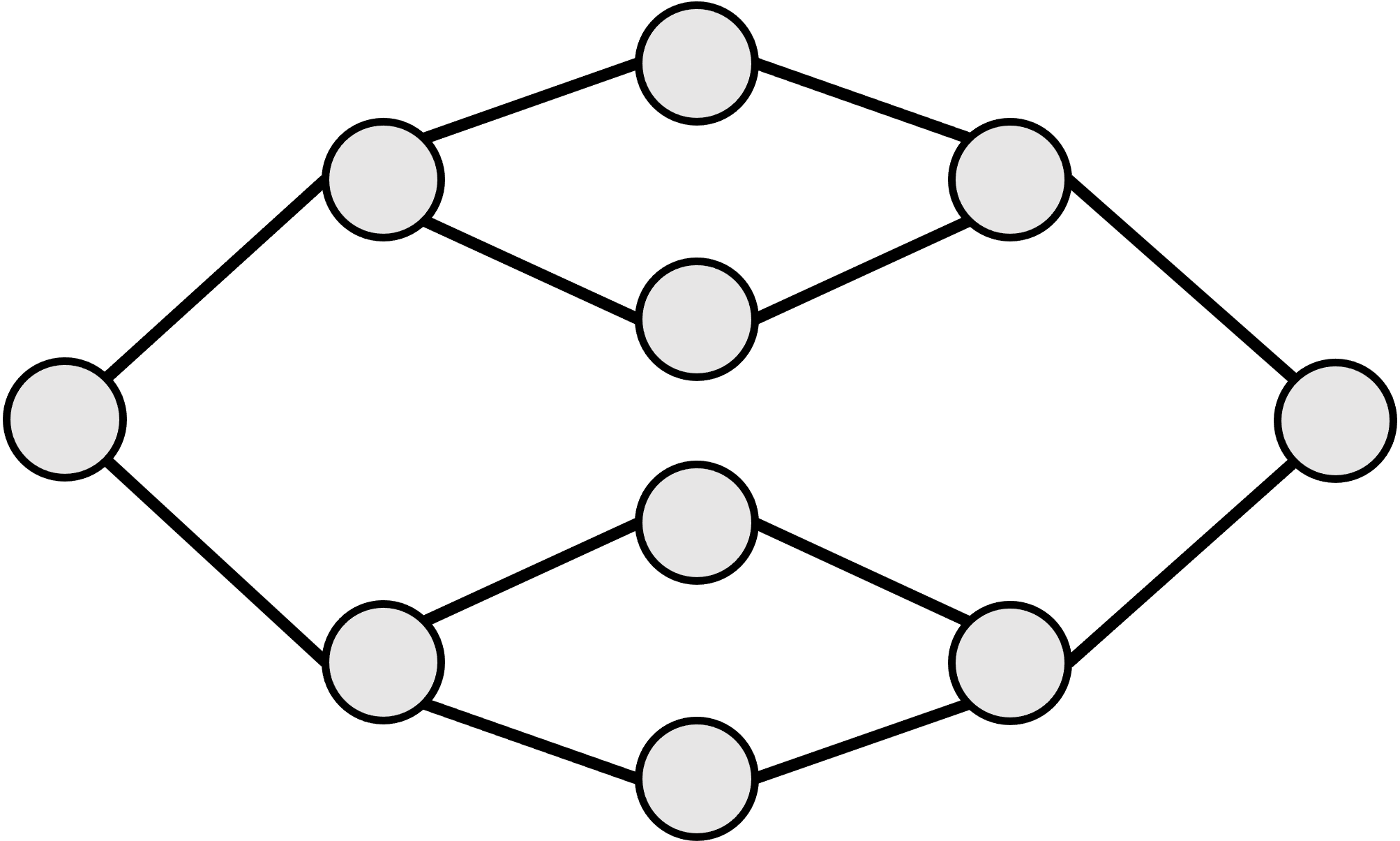}
    }
    \hfill
    \end{minipage}
    \caption{Examples of graph-based circuit interaction graphs.}
    \label{fig:circuit-graphs}
\end{figure}%

\subsection{Architectures}
One of the biggest limitations on a quantum architecture is simply the number of qubits available for execution, which constrains many currently feasible programs. Therefore, for different strategies across circuit sizes, we test circuits on regular architectures that are just large enough for the circuit in question.  We use a rectangle grid-based mesh as our architecture, with dimensions of $\lceil\sqrt{n}\rceil \times \frac{n}{\lceil\sqrt{n}\rceil}$ where $n$ is the number of qubits in the circuit. Each qubit is connected to four other qubits directly adjacent to it except along the border of the architecture. This construction allows for testing the scalability of our methods and test performance in the edge cases of high architectural usage.

We also examine our methods on the 65-qubit IBM Ithaca heavy-hex topology \cite{ibm_quantum_ibm_nodate} and a 65-qubit ring architecture. This gives us a sense of how well our methods work on devices with more constrained connectivity.  

\subsubsection{Gate and Coherence Times Statistics}
We have shown gate times found via optimal control in Section \ref{sec:gate_durations}.  We use the target fidelities for these gates as their success rates.  Single-qudit gates (regardless of Hilbert dimension) are optimized to a success rate of 99.9\% and two-qudit gates to 99\% when simulated in an ideal scenario \textit{without noise}.  The product of the success rate of every gate in the circuit is the Expected Probability of Success (EPS) with respect to gate execution.

To account for increasing noise in higher dimensional systems we consider their coherence times ($T_1$), after which the qubits are unable to maintain their elevated energy state.  For a $d$-level system, the coherence time is estimated to be $\nicefrac{1}{d-1}$ of the $T_1$ time for the original two-level system \cite{blok_quantum_2021}. Critically, because gate durations get longer with corresponding lower $T_1$ times we capture the increasing susceptibility to error in a mixed-radix system. We  use a 163.5 $\mu$s qubit $T_1$ time in our architecture, which gives a 54.5 $\mu$s worst case $T_1$ time when we are in a ququart state. \cite{ibm_quantum_ibm_nodate}.  The product of the probabilities of no decoherence for each qubit, $e^{-t_{qb}/T_{1qb} - t_{qd}/T_{1qd}}$, is the EPS with respect to coherence time. Here $t_{qb}$ is time spent in the qubit state, and $t_{qd}$ is the time spent in the ququart state.

The product of these two statistics gives the overall EPS for the entirety of the circuit, and gives an upper bound for the amount of error found in the circuit.  It is the worst case for both gate fidelity and coherence time as it assumes each qubit will be measured and used for the entire duration of the circuit, and uses the worst case coherence time ratio from qubits to qudits.  However, it does not take into account the effects of more complicated errors such as crosstalk.

\subsection{Baselines}
We compare against two different baseline strategies, representing the extremes of compilation for qubits on a ququart-supported device.

\subsubsection*{Qubit-Only Compilation}
One extreme is to never encode any \\ ququarts. In this case, we use EQM but never allow exploration of the second encoded position.  These strategies are comparable to other compilation pipelines described in \cite{murali_noise-adaptive_2019}. 

\subsubsection*{Full Ququart Pairing with Encoding and Decoding (FQ)}
There has been discussion of the potential of fast internal ququart operations and use ququarts for general qubit circuit compilation \cite{baker_efficient_2020}.  However, without partial ququart operations, which were assumed to be as expensive as any multi-qudit operation, any external operation required decoding the two qubits, performing an operation, and re-encoding. A drawback of this strategy is that extra space is always required, since a decoding operation requires an ancilla bit to decode into.

We are unaware of an automatic compilation pipeline constructed for this strategy.  We use an EQM strategy which allocates extra decoding space next to fully encoded ququarts as the mapping strategy.  For routing, we use a similar strategy, but only at the qudit level. This means that we only use full qudit SWAP gates to ensure two qudits are adjacent to on another. Additionally, we must route the empty decoding ancilla next to the neighboring qubits so that we may perform decompression.

\begin{figure*}
    \centering
    \centering
    \scalebox{0.85}{
    \includegraphics[width=\linewidth]{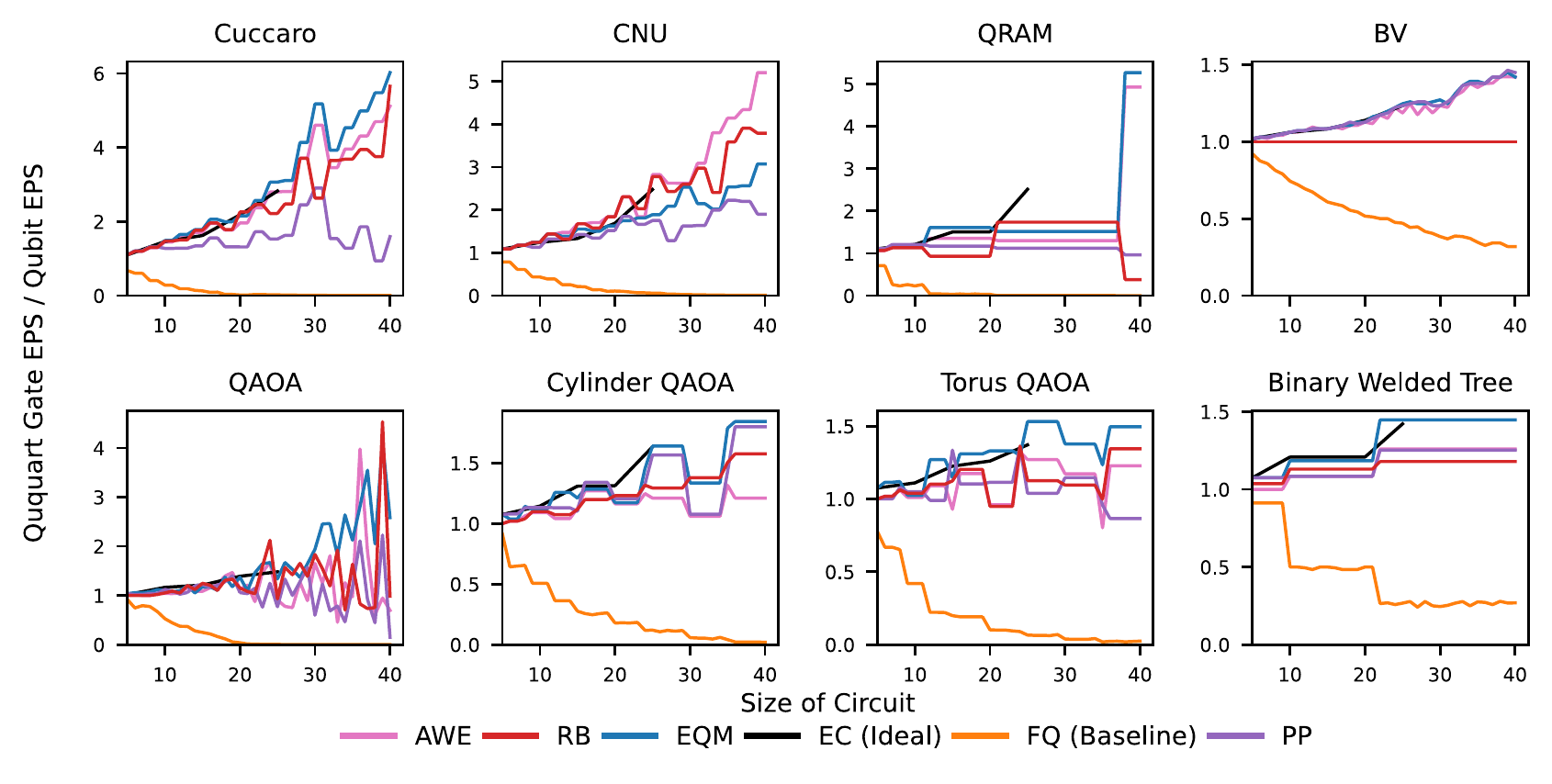}
    }
    \caption{Expected Gate Probability of Success for each benchmark.  Each color represents a a different compilation strategy, where the black line is the exhaustive solution developed previously, and is the goal. FQ is the previous baseline for generalized ququart computation.}
    \label{fig:crit-path-fidelity}
\end{figure*}

\begin{figure}
    \centering
    \scalebox{0.85}{
    \includegraphics[width=\linewidth]{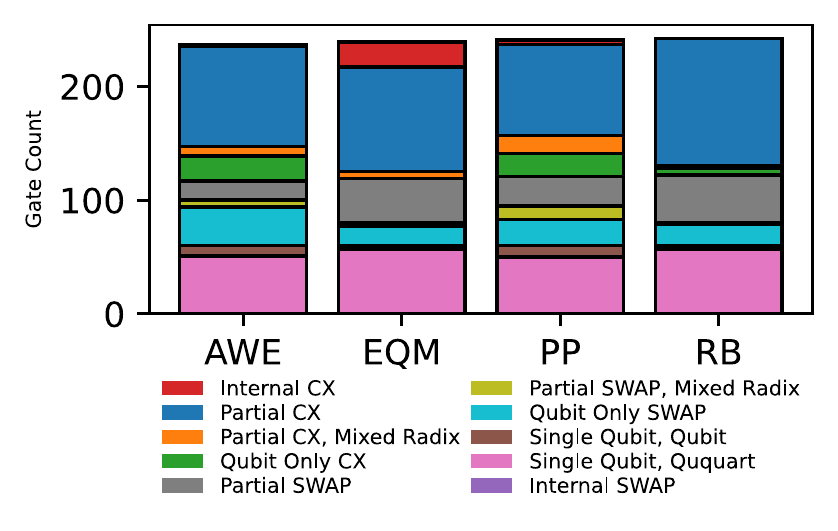}
    }
    \caption{The distribution of gate types for different pairing strategies for a 30 qubit Torus QAOA circuit.  Each color represents a different ``style'' of gates.  In particular the darker blue represents CX gates between two ququarts, and the red represents an internal CX gate within a single ququart.}
    \label{fig:gate-distribution}
\end{figure}

\begin{figure}
    \centering
    \scalebox{0.85}{
    \includegraphics[width=\linewidth]{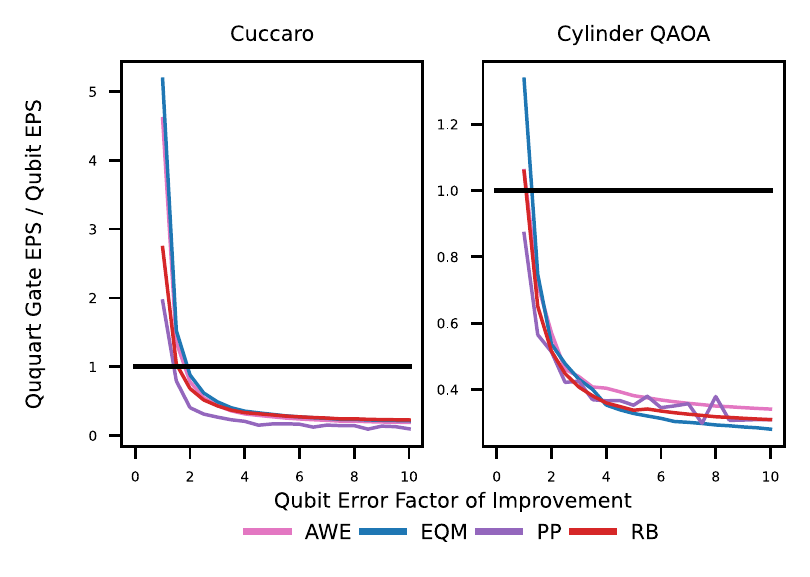}
    }
    \caption{Gate Expected Probability of Success as the qubit gate error rate increases and the ququart error gate rate stays constant.  The black line represents the crossover point where the error of the qubit only compilation is greater than ququart compilation.}
    \label{fig:gate-sensitivity}
\end{figure}

\subsection{Benchmarking}

We use a number of different types of circuits to explore many use cases to test the viability of each compression strategy.  One set of circuits we explore have localized groups of qubits that interact together.  They have no potentially random operations that disrupt the grouping of operations. These include the Cuccaro Adder \cite{cuccaro_new_2004}, Generalized Toffoli (CNU) \cite{barenco_elementary_1995}, Quantum RAM (QRAM) \cite{gokhale_quantum_2020} and Bernstein Vazarani (BV) \cite{bernstein_quantum_1997}.

We are also interested in certain graph-based interaction structures. We use a QAOA construction \cite{farhi_quantum_2014} that accepts a graph where each node represents a qubit and an edge represents an interaction. For each edge, in a random order, we perform a CX, a Z gate, and another CX gate.  These circuits are not necessarily used in practice, but allow for the examination of particular interaction structures which may be relevant in optimization problems.  We examine a random graph with edge density 30\%, a cylindrical graph shown in Figure \ref{fig:circuit-graphs}(a), a similar torus structure (Figure \ref{fig:circuit-graphs}(b)) and a binary welded tree (Figure \ref{fig:circuit-graphs}(c)).

This set of circuits were selected to be representative of different interaction graphs, and to explore the effects on mostly parallel circuits (Generalized Toffoli) versus mostly serial circuits (Cuccaro Adder and QRAM). Each of these circuits is tested across a range of different sizes to examine how these different strategies scale as the size of the circuit increases.

These compilers were constructed on top of the Qiskit library, version 0.18.3 \cite{anis_qiskit_2021}, on Python 3.9 \cite{van_rossum_python_2009}. The benchmarks were run on a machine with Intel(R) Xeon(R) Silver 4110 2.10GHz, 132 GB of RAM, on Ubuntu 16.04 LTS.

%% file: paper-text/08-results.tex

\begin{figure*}
    \centering
    \scalebox{0.85}{
    \includegraphics[width=\linewidth]{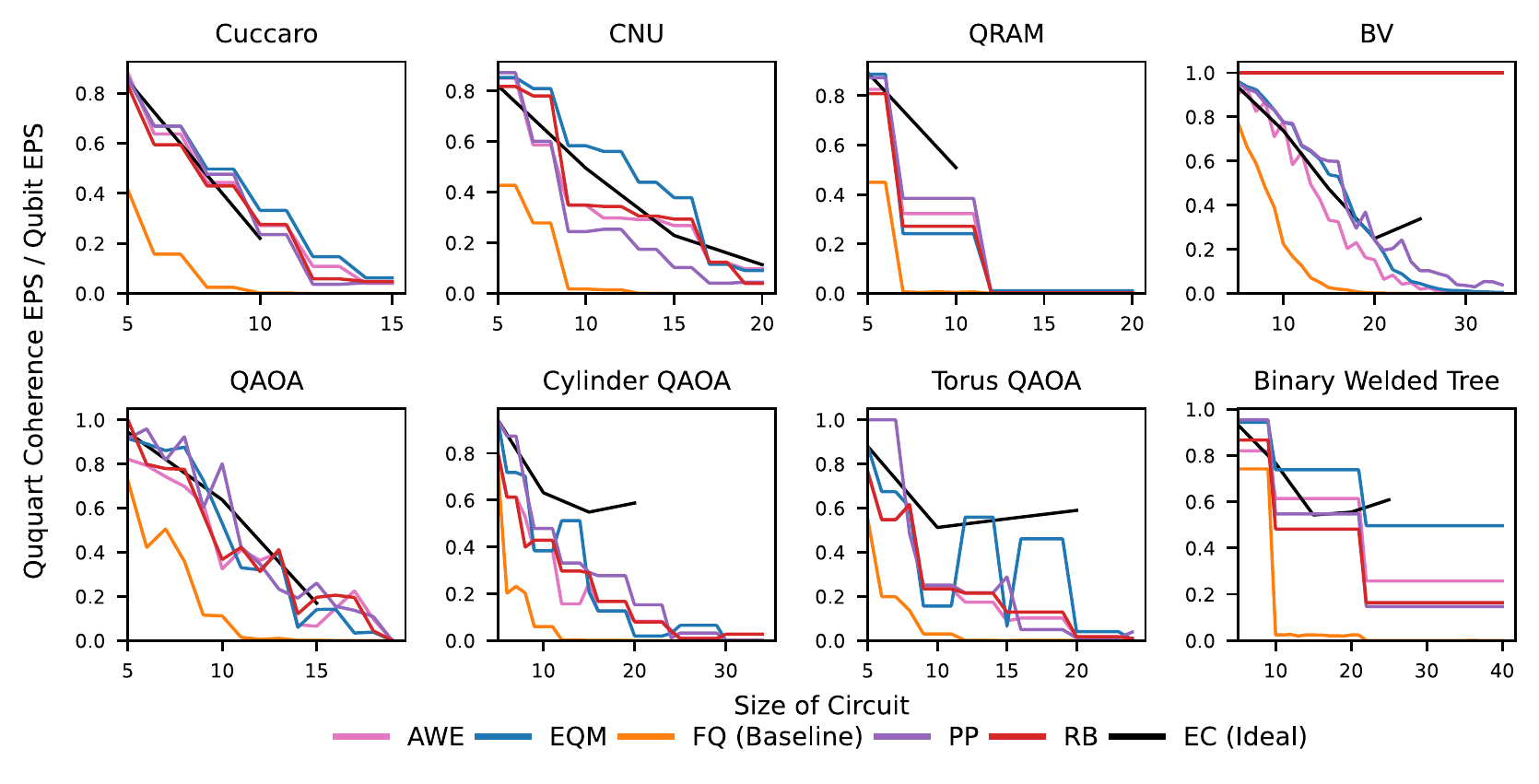}
    }
    \caption{Expected Coherence Probability of Success for each benchmark.  Each color represents a a different compilation strategy, where the black line is the exhaustive solution, and in theory ideal, developed previously. EC line stops short for computational reasons, requiring many more classical resources. FQ is the previous baseline for generalized ququart computation.}
    \label{fig:time-fidelity}
\end{figure*}

\begin{figure}
    \centering
    \scalebox{0.85}{
    \includegraphics[width=\linewidth]{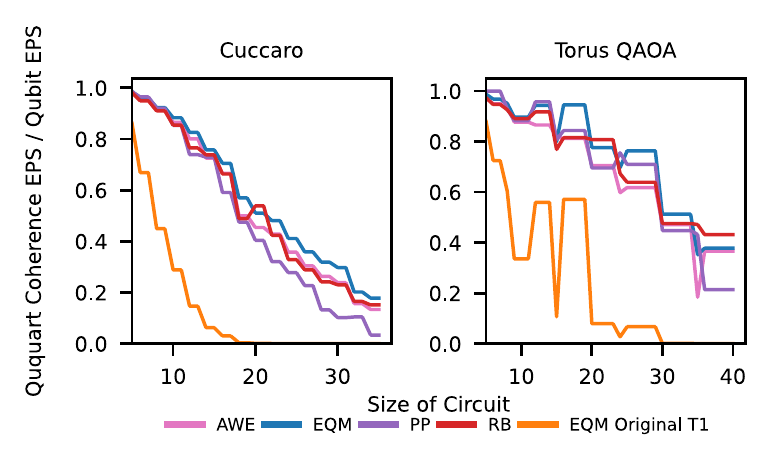}
    }
    \caption{Expected Coherence Probability of Success for Cuccaro and Torus QAOA with 10x better ququart and qubit $T_1$ times.}
    \label{fig:10x-better-t1}
\end{figure}

\begin{figure}
    \centering
    \scalebox{0.85}{
    \includegraphics[width=\linewidth]{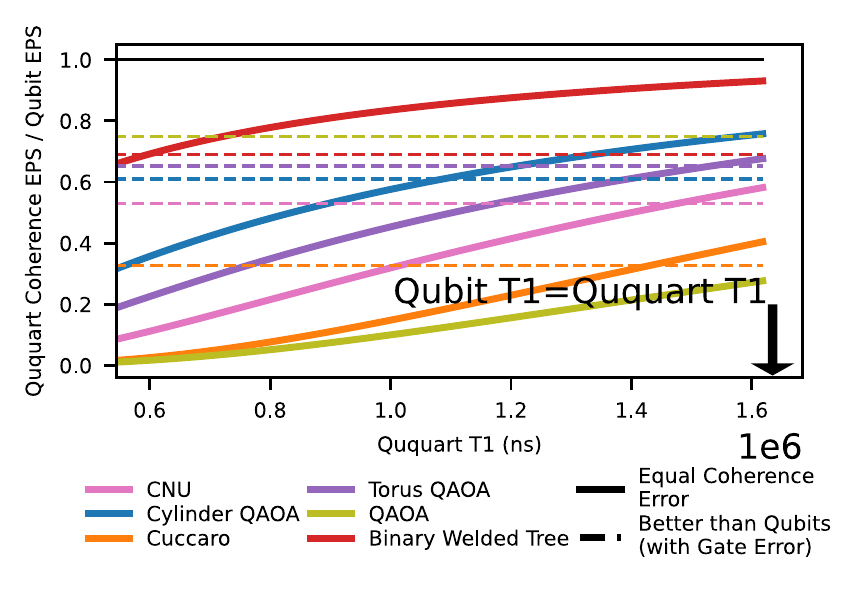}
    }
    \caption{Expected Coherence Probability of Success for several 25 qubit benchmarks with 10x better ququart and qubit $T_1$ times as the ququart time increases $T_1$ from 1/3 the qubit $T_1$ time to the entire qubit $T_1$ time.  Each dashed line represents the point where the coherence success rate no longer outweighs the gate success rate gains for the corresponding circuit.}
    \label{fig:increasing-ququart-time}
\end{figure}

\section{Results}

Two benefits of encoding into ququarts are the decrease in required gates due to reduced communication requirements and the transformation of some two-qubit gates into much faster single-ququart gates.  In Figure \ref{fig:crit-path-fidelity}, we examine how the  gate expected probability of success (EPS) compares to qubit-only compilation for each compression strategy on the same architecture.

Primarily, FQ (orange line, baseline) is consistently worse than our qubit-only baseline.  This is expected. Every out-of-pair operation requires more communication, plus the decode and encode steps.  As a result, we actually see increases in the number of gates, as well as a decrease in the overall circuit gate fidelity.

Certain compression strategies are more advantageous than others on different circuit constructions.  In the more regularly structured circuits, namely CNU and Cuccarro adder, we see the greatest gains in circuit fidelity due to gates from EQM (blue) and RB (red) strategies, with improvements over 50\% for both.  In fact, these gains match or exceed the EC (black, ideal) case, which requires much more classical computation and is impractical for even moderately sized programs. These circuits have focused interaction on varying sets of qubits over time, as seen in Figure \ref{fig:ring-examples}.  The circuit interaction graphs are very easy to flatten into a line by compressing qubits within cycles, eliminating the need for communication. Then only CX operations are needed.

However, the RB strategy is less consistent for BV and QRAM circuits.  For BV, this makes sense; there are no cycles to examine in the interaction graph, so no compressions are made.  But, QRAM has many cycles.  In this case, because the cycles share edges rather than nodes, a compression on one cycle may adversely affect others. Predicting interactions between the cycles is more computationally difficult and was not explored.  However, EQM can more closely mimic the interaction graph's connectivity on the architecture, reducing the number of SWAP gates and increasing internal CX interactions.

In the graph-based circuits, where each edge is weighted similarly, no method clearly wins over the other strategies and we only find up to 20\% improvements in gate success rate.  The most consistent performer is still EQM, which almost never drops below the corresponding qubit compilation success rate.  On the other hand, strategies that prioritize communication, such as AWE and PP, are much more inconsistent. These strategies group qubits together that reduce the distance to interact with more qubits.  While this might seem to decrease communication, it does not. Instead, qubits are constantly shifted into positions where there is not much locality to exploit.  That is, highly-interacting qubits are not necessarily placed close to one another, increasing the chance that communication will be required. These methods significantly reduce the number of internal ququart operations and increase serialization of communication due to both qubits in a pair being required independently for communication. This is shown in Figure \ref{fig:gate-distribution}.  The overall numbers of gates are similar between EQM and AWE, but EQM uses significantly more internal CX gates (red bars). AWE and PP tend to use more SWAP gates (grey, green, and cyan bars) and many more partial CX operations (orange and blue bars). EQM enables more success rate improvement by prioritizing higher fidelity multi-qubit operations rather before prioritizing communication reduction.

While communication reduction is important, it is difficult to predict prior to compressions being made and recompilation, and it is more crucial to find the compressions that will increase internal CX count early in the circuit with a general sense of global scope, and leave the router to dynamically adapt to changing communication costs.

\paragraph{Sensitivity to Better Qubit Error}
It may not be the case that ququart and qubit gates have identical error.  As mentioned previously, ququart gates are more difficult to control and may have lower fidelity in a real system.  In Figure \ref{fig:gate-sensitivity}, we demonstrate how strategies react to higher fidelity qubit-only gates for the Cuccaro circuit and for the cylinder QAOA.  The strategies maintain their relationship to each other, but, as expected, see diminishing returns as qubit error improves.  However, there is some variability in the pairing methods as the qubit error rate improves, as they attempt to preference qubit operations as they become less error prone. Even in these cases, the underlying architecture may be \textit{space limited} meaning we require larger hardware than required by a mixed-radix strategy.

\subsection{Error Due to Circuit Duration}

Ququart compressions have a downside. Each compression induces serialization and requires the use of the longer partial ququart gate times.  In Figure \ref{fig:time-fidelity}, we explore how each compression strategy affects the error due to increasing circuit duration. As the circuit duration increases, a qudit is more likely to decohere due to the significantly worse $T_1$ time. The probability of an entire system maintaining coherence for a mixed-radix circuit is described asthe product of $e^{-t_{qb}/T_{1qb} - t_{qd}/T_{1qd}}$ for each qubit. $t_{qb}$ is the time spent in the qubit state and $t_{qd}$ is the time spent in the ququart state, and $T_{1x}$ is the $T_1$ coherence time for the qudit in the noted state. We first notice that we significantly improve upon the time incurred by FQ; all other compression strategies are able to more effectively mitigate circuit duration increases. Partial SWAP operations are faster than full qudit SWAP operations and prevent extraneous communication. We also find that EQM-based compression generally leads to the best coherence probability of success. Furthermore we observe that, in some cases, the highest gate probability of success does not always match the best coherence probability of success.  This is due to the fact that some single-qubit and multi-qubit gates can no longer be performed in parallel due to compression, significantly contributing to the circuit duration.  We also notice that we are able to mostly match the duration found through exhaustive search of critical path success rates.

However, a total success rate is the success rate via gate fidelity product times the coherence time success rate, and at current $T_1$ times (listed in Section \ref{sec:evaluation-methods}) decoherence error outweighs the benefits of success rate increases.  We examine the coherence probability of success for a Cuccaro circuit and Torus QAOA in Figure \ref{fig:10x-better-t1}, with a 10x better T1 time for both qubits and ququarts.  While the margin between qubit and ququart circuits improves, it will still outweigh the gate success.

\begin{figure}
    \centering
    \scalebox{0.85}{
    \includegraphics[width=\linewidth]{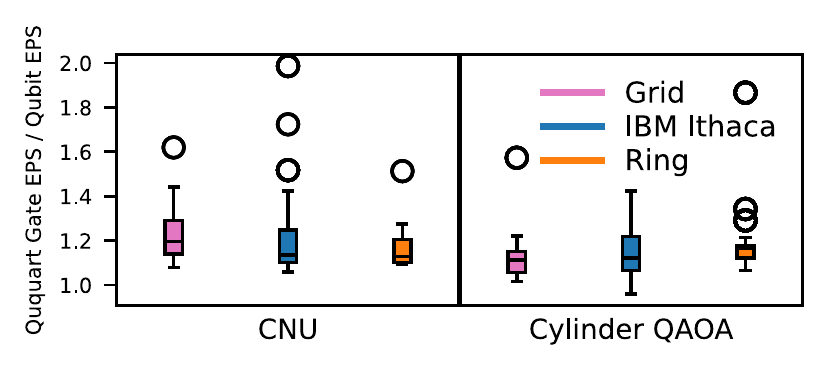}
    }
    \caption{Ranges of gate based probability of success for CNU and Cylinder QAOA on three different architectural topologies.  This is the combined set of ratios of improvement for circuits sizes 5 to 40.}
    \label{fig:architecture-variation}
\end{figure}

However, the $1:3$ ratio of $T_1$ times used in this work is a worst case scenario. Two qubits encoded in a ququart may not be in the maximum $\ket{3}$ state at all times, and will not be subject to the same loss in $T_1$ time at all points.  Additionally, physically realized qutrit devices have seen $T_1$ relationships that are better than the expected $\nicefrac{1}{2}$ reduction \cite{blok_quantum_2021} and can be specifically designed to enhance the expected decay rates for different energy levels.  Using the circuit durations found here, we plot the change in success rate due to circuit duration as the ratio of $T_1$ time changes in Figure \ref{fig:increasing-ququart-time}.  As the $T_1$ for ququarts increases, we find that there is a point, the dashed lines, before the $T_1$ times are equal where the total success rate is improved with ququarts.  In these instances, ququarts would be expected to perform better than qubits.  While it is difficult to avoid an increase in circuit duration, with enough gains through internal CX gates and SWAP count reduction, encoding qubits into ququarts has the ability to extend what can be computed on a device and perform with better expected probability of success. 

\subsection{Computation on Lower Connectivity}

Most quantum architectures, especially superconducting devices, have much lower connectivity than the grid architecture we assumed.  As described in Section \ref{sec:evaluation-methods}, we also test our methods on the IBM Ithaca topology and a similarly sized ring-based topology.  We describe the range of critical path fidelity improvement for some circuits in Figure \ref{fig:architecture-variation}. The patterns shown are consistent across all benchmarks.  We do not see any significant difference in performance between architectures.  This is expected. We use the similar routing steps for both qubits and ququarts, the main difference being that the connectivity is slightly expanded for the ququart routing. Our methods can successfully adapt to different structures with similar effects for each.

%% file: paper-text/09-conclusion.tex
\section{Discussion and Conclusion}

Architectural size is a huge limitation on useful computation on quantum computers, and many architectural models for quantum architectures have access to higher level states.  These higher level states can be used to compress quantum information.  In the case of ququarts, two qubits can be fit into a single physical unit in a process we call \textit{compression}.  There are difficulties in working with these more complex objects such as shorter coherence times, and longer gate times work against each other to make ququart compilation challenging.

In this work, we explicitly realize a gate set for mixed qubit-ququart systems including partial operations, without the need to encode and decode ququarts. With a gate set designed specifically for ququart computation and communication, we design and evaluate a compiler to mitigate time-intensive ququart-ququart interactions and the increased gate-based communication required to interact two ququarts in past models. Our compiler also exploits higher qubit connectivity and reduces the count of needed physical units to execute programs requiring more qubits than available.  By mitigating this cost, we explore the best way to take advantage of  faster operations, such as internal CX gates, enabled by encoding multiple qubits in the same physical unit.  We develop several strategies that emphasize the need to prioritize these fast interactions and exploit the locality of certain circuit structures for better gate expected probability of success, and lower increases in circuit duration.

We demonstrate the potential of ququart logic as a valuable tool in making quantum devices more useful  in the future.  We show the potential for doubling the number of qubits available for execution, and guide future device engineering and architectural development to prioritize these higher energy states.